\documentclass[aps,prl,twocolumn,groupedaddress]{revtex4}
\usepackage{ graphicx}

\begin{document}

\title{Condensed states of a semiflexible copolymer in poor solvent:
Figures of eight and discrete size torii}

\author{Hern\'andez-Zapata, E.}
\author{Cooke I. R.}
\author{Williams D. R. M.}
\affiliation{Department of Applied Mathematics, ANU, Australia}

\date{\today}

\begin{abstract}
We examine the condensed states of a simple semiflexible copolymer in
which there are two monomer types that are immiscible with each other
and with the solvent.  Although this is similar to the well known
problem of collapse for a semiflexible homopolymer we find that it
gives rise to a much richer variety of condensed states.  We predict
the existence of these states using simple analytic arguments and also
observe them directly using Brownian dynamics simulations.
\end{abstract}

\pacs{0001}

\maketitle 

One of the most fascinating physical properties of DNA is its collapse
from a diffuse coil to a compact toroidal structure under the addition
of condensing agents to the solvent.  This phenomenon is not only of
biotechnological importance \cite{fang_hoh:1999,blessing_etal:1998}
but also serves to illustrate a general principle of competition
between bending and solvent energy that is common to the collapse of
all semiflexible polymers including other biopolymers, synthetic
polymers and also carbon nanotubes \cite{martel_etal:1999}.  Although
much progress has already been made in understanding the kinetics of
folding
\cite{sakaue_yoshikawa:2002,schnurr_etal:2000,montesi_etal:2003} as
well as the internal packing arrangements of semiflexible polymer
torii
\cite{pereira_williams:2000,hud_etal:1995,hud_downing:2001}
much remains that is not understood.  Our intention in this paper is
to highlight the possibilities that arise when one extends the problem
of semiflexible polymer collapse in poor solvent to a copolymer.
Inspired by the fact that DNA torii occur as a consequence of a
competition between solvent and bending energies we study copolymers
with two block types A and B that are immiscible with each other and
with the solvent.  We begin with an analytic treatment of the problem
which leads us to predict several unexpected stable geometries.  Using
Brownian dynamics simulations we demonstrate that such geometries can
self assemble via collapse of a single semiflexible copolymer and that
their relative energies are in agreement with the theory.

\section{Theory}
As a starting point for our analytic theory we review some basic
results for a semi-flexible homopolymer.  The packing arrangement of a
semiflexible polymer that optimizes bending within the torus is
certainly not a trivial matter and has been treated elsewhere
\cite{pereira_williams:2000}.  For simplicity we will
assume a simple garden hose or circular packing.  We will also assume
that the chain enrolls several turns in such a way that it is thick
enough to give a well defined toroidal surface, but it is thin enough
so that the minor radius of the toroid, $r$, is much less than the
major radius, $R$. For such a thin toroid, every section of the chain
has approximately the same curvature. Using two theorems of Pappus,
the surface area and volume of the toroid are: $A = 4 \pi^{2} R r$ and
$V = 2 \pi^{2} R r^2$.  Therefore, the free energy of the chain can be
written as:
\begin{equation}
\Delta F = \frac{\kappa L}{2 R^2} + 4 \pi^{2} \gamma R r
\label{eqn1}
\end{equation}
Here the first term represents the bending energy while the second
term is the surface energy, $L = N l$ is the total length of the
polymer along the chain, $N$ is the number of repeating units, $l$ is
the monomer length, $\kappa$ is the bending constant, and $\gamma$
is the surface tension (which is in this model the parameter that
characterizes the polymer-solvent interaction).

Using the volume constraint $V=La^2$ ( $a^2$ is the cross section of
the chain ) the minor radius of the toroid can be written as, $r =
\frac{a}{\sqrt{2} \pi} \sqrt{\frac{L}{R}}$ and equation \ref{eqn1}
transforms to:
\begin{equation}
\Delta F = \frac{\kappa L}{2 R^2} + 2 \sqrt{2} \pi
\gamma a \sqrt{L R}
\label{eqn2}
\end{equation}
If the free energy is minimized with respect to the major radius of
the toroid, the following dependence is obtained:
\begin{equation}
\frac{R}{a} = \left( \frac{\alpha}{\sqrt{2} \pi} \right)^{\frac{2}{5}}
b^{\frac{1}{5}} N^{\frac{1}{5}}
\label{eqn3}
\end{equation}
Here we have used the dimensionless parameters $\alpha \equiv
\frac{\kappa}{\gamma a^{3}}$ and $b \equiv \frac{l}{a}$.

Consider now a more complex problem: a semi-flexible copolymer formed
by a sequence of two different immiscible blocks (which we call
''block A'' and ''block B''). For simplicity, an equal size for both
blocks will be assumed. If the repeating unit along the chain (that
is, the sequence of a block A followed by a block B) is called a
diblock, then, for a copolymer, $l$ will denote the length of a
diblock while $N$ will denote the total number of diblocks along the
chain. The repulsive interaction between blocks A and B is assumed to
be dominant over all other interactions such that it may be modeled by
imposing the restriction that a block A cannot have lateral contact
with a block B. According to this restriction, if the copolymer still
forms a toroid in poor solvent conditions, the ratio of the toroid
perimeter, $2 \pi R$, and the diblock length, $l$, should be
necessarily a whole number; that is,
\begin{equation}
R = m \frac{l}{2 \pi},\verb+ + m= 1, 2, 3, \verb+...+
\label{eqn4}
\end{equation}
Here the number $m$ denotes the number of diblocks per turn in the
toroid.  The free energy of the toroid still has the same form as in
equation \ref{eqn2} and since block types A and B will allways have
equal surface areas within our model we can make the interpretation
$\gamma \equiv \frac{(\gamma_A + \gamma_B)}{2}$, where $\gamma$ is the
average surface tension of block types A and B.  For simplicity we
also assume that $\kappa_A \equiv \kappa_B$ so that only a single
$\kappa$ is necessary. We should note however, that the value of the
radius $R$ is now restricted to be chosen from a set of discrete
values, given by equation \ref{eqn4}.  Hence the free energy itself
varies in a discrete way:
\begin{equation}
\left( \Delta F \right)_m = \frac{2 \pi^{2} \kappa}{l m^{2}} N + 2
\sqrt{\pi m} \gamma a l N^{\frac{1}{2}}
\label{eqn5}
\end{equation}
The transition from a toroid with $m+1$ diblocks per turn, to a toroid
with $m$ diblocks per turn occurs when the difference in free energy,
$ \left( \Delta F \right)_{m+1} - \left( \Delta F \right)_m$, changes
from negative to positive.  If we define the parameters $\Upsilon$ and
$f_m$ as in equations \ref{eqn6a} and \ref{eqn6} then this will occur when $\Upsilon
\leq f_m$.
\begin{eqnarray}
\label{eqn6a}
\Upsilon = \frac{\pi^{\frac{3}{2}}
\alpha N^{\frac{1}{2}}}{b^2}  = \frac{\pi^{\frac{3}{2}}
 \kappa N^{\frac{1}{2}}}{\gamma a l^{2}} \\
f_m = \left( \sqrt{m+1} -
\sqrt{m} \right) \left( \frac{1}{m^2} - \frac{1}{\left( m+1
\right)^{2}} \right)^{-1} 
\label{eqn6}
\end{eqnarray}
In general, for $m > 1$ , the toroid with $m$ diblocks per turn will
be stable if the condition \ref{eqn7} is satisfied.
\begin{equation}
f_{m-1} \leq \Upsilon \leq f_m
\label{eqn7}
\end{equation}
When $m \gg 1$ the discrete jumps in $R$ should be small compared to
$R$, so that we should recover the behavior of a homopolymer, given by
equation \ref{eqn3}. Note that when $m$ is big, we can approximate
$f_m \approx \frac{m^{\frac{5}{2}}}{4}$ and the transition points
occur when $\frac{m^{\frac{5}{2}}}{4} \approx \Upsilon$. Therefore the
radius of the toroid in the transition, given by the restriction
\ref{eqn4}, becomes $\frac{R}{a} = m \frac{l}{2 \pi} \approx \left(
\frac{\alpha}{\sqrt{2} \pi} \right)^{\frac{2}{5}} b^{\frac{1}{5}}
N^{\frac{1}{5}}$, coinciding with equation \ref{eqn3} as expected.

However, when the parameter $\Upsilon$ is small (that is, when the
surface energy dominates over the bending energy) the radius of the
toroid will reach its lower limit, $R = \frac{l}{2 \pi}$,
corresponding to $m=1$ diblocks per turn.  This lower limit to $R$
also implies a lower limit to the surface area within the toroidal
geometry.  Thus we expect that as $\Upsilon$ becomes still smaller and
surface energy even more important the polymer must explore
alternative conformations to the simple torus.

\begin{figure}
\label{fig8diag}
\includegraphics{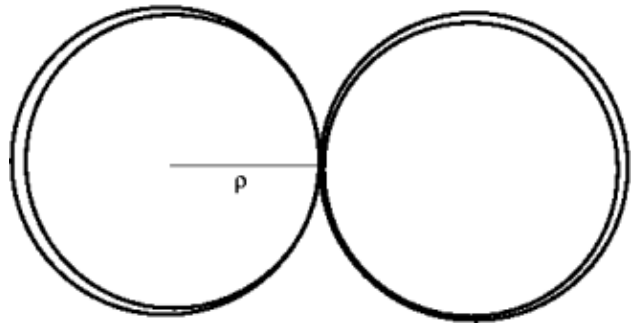}
\caption{Schematic diagram of a polymer in the figure 8 configuration
with $n=2$}
\end{figure}

One possible conformation, for small values of the parameter
$\Upsilon$, would be a figure 8 conformation, such as that shown in
figure \ref{fig8diag}. In this conformation a block A enrolls a whole
number of turns, $n$, around a circle of radius $\rho$ in the left
side, the subsequent block B enrolls $n$ turns around a circle of the
same radius in the right side. and so on. The free energy of the
copolymer in this figure 8 conformation is approximately equal to the
free energy of two separated toroids of radius $\rho$, but each of
them with a total chain length $\frac{L}{2}$ instead of
$L$. Therefore, the expression for the free energy, $\Delta F_8$,
corresponding to the figure 8 conformation, would be:
\begin{equation}
\Delta F_8 = \frac{\kappa L}{2 \rho^{2}} + 4 \pi
\gamma a \sqrt{L \rho}
\label{fr8}
\end{equation}
Now the restriction of no lateral contact between blocks A and B is
imposed. Since each block (of length $\frac{l}{2}$) enrolls $n$ turns
(with $n$ a whole number) around the circle of radius $\rho$, this
leads to:
\begin{equation}
\rho = \frac{l}{4 \pi n} \verb+ +n =1, 2, 3,\verb+...+
\label{eqn8}
\end{equation}
Lets forget momentarily this last restriction and study the associated
problem of a homopolymer that somehow (for example, by imposing some
unspecified restrictions) has acquired a figure 8 conformation with
free energy given by equation \ref{fr8}. The equilibrium radius,
$\rho$, of both sides of the figure 8 would be: $\rho = \left(
\frac{\kappa}{2 \pi \gamma a} \right)^{\frac{2}{5}} L^{\frac{1}{5}}$.
Substituting this result in equation \ref{fr8} we obtain a free energy
$\Delta F_8 = \frac{5}{2} \left( 2 \pi a \gamma \right)^{\frac{4}{5}}
\kappa^{\frac{1}{5}} L^{\frac{3}{5}}$. Comparing with the analogous
free energy of the toroid conformation of a homopolymer, $\Delta F =
\frac{5}{2} \left( \sqrt{2} \pi a \gamma \right)^{\frac{4}{5}}
\kappa^{\frac{1}{5}} L^{\frac{3}{5}}$, we can readily see that $\Delta
F_8$ is always bigger than $\Delta F$ by a factor of
$2^{\frac{2}{5}}$. Therefore, in the case of a homopolymer, where the
value of the radius is not restricted, the toroid conformation is
always more stable than the figure 8 conformation. This is not
surprising, since there is obviously an increase in bending when going
from a torus to a figure 8.  As noted previously, however, in the case
of a semi-flexible copolymer, for small values of the parameter
$\Upsilon$, i.e. when the surface energy dominates over the bending
energy, the toroid major radius is not allowed to be smaller than $R =
\frac{l}{2 \pi}$, while the radius, $\rho$, of both sides of the
figure 8 conformation can still be as small as needed, as long as they
satisfy the restriction \ref{eqn8}.  Thus for small $\Upsilon$ it is
possible for the figure 8 to be more stable than the torus.

The free energy of a figure 8 conformation with $n$ turns per diblock
is obtained by substituting equation \ref{eqn8} into equation
\ref{fr8}, leading to:
\begin{equation}
\left( \Delta F_8 \right)_n = \frac{8 \pi^{2} N n^2 \kappa}{l} +
\frac{2 \sqrt{\pi} \gamma a l N^{\frac{1}{2}}}{\sqrt{n}}
\label{eqn9}
\end{equation}
A figure 8 conformation would be more stable than a toroid
conformation if the free energy of the toroid with one diblock per
turn, $\left( \Delta F \right)_1$, is bigger than the free energy of a
figure 8 conformation, $\left( \Delta F_8 \right)_n$ for some value of
the number of turns per diblock, $n$. Using equations \ref{eqn5} and
\ref{eqn9}, this condition leads to:
\begin{eqnarray}
\left( \Delta F \right)_1 - \left( \Delta F_8 \right)_n
= & \nonumber \\
= \frac{2 \pi^{2} \kappa N}{l} \left( 1-4 n^2
\right) + 2 \sqrt{\pi}
\gamma a l N^{\frac{1}{2}} \left( 1-
\frac{1}{\sqrt{n}} \right) \geq 0 \nonumber
\\
\Longrightarrow g_n \equiv \left( 1-
\frac{1}{\sqrt{n}} \right)
\frac{1}{\left( 4 n^2 -1 \right)} \ge \Upsilon
\label{eqn10}
\end{eqnarray}
Note that $g_1 = 0$ and hence there cannot be a transition from a
toroid conformation to a figure 8 conformation with just one turn per
diblock, $n = 1$. However, for values of the parameter $\Upsilon$
smaller than $g_2 \approx 0.0195$, a transition to a figure 8
conformation with two turns per diblock would be favourable.

Finally, transitions between figure 8 conformations with different
values of $n$ can also be studied by analyzing the difference between
their free energies, $\left( \Delta F_8 \right)_n - \left( \Delta F_8
\right)_{n+1}$. When this difference becomes bigger than zero, the
conformation with $n+1$ turns per diblock becomes more stable than the
conformation with just $n$ turns per diblock. It is easy to see that
\begin{eqnarray}
\left( \Delta F_8 \right)_n - \left( \Delta F_8 \right)_{n+1} \geq 0 &
\nonumber \\ \Longrightarrow h_n \equiv \frac{1}{4} \left(
\frac{1}{\sqrt{n}} - \frac{1}{\sqrt{n+1}} \right) \frac{1}{\left[
\left( n+1 \right) ^2 -n^2 \right]} \geq \Upsilon
\end{eqnarray}

Summarizing the previous discussion the following regimes are
expected:

(i) In the case in which inequality \ref{eqn7} is valid, a toroidal
conformation with $m$ diblocks per turn (for $m > 1$).  The radius of
the toroid would be given by equation \ref{eqn4}.

(ii) If $g_2 \leq \Upsilon \leq f_1$, then a toroidal conformation
with one diblock per turn would be expected. The toroid radius would
be given by $R = \frac{l}{2 \pi}$.

(iii) If $h_2 \leq \Upsilon \leq g_2$, then a figure 8 conformation
with one turn per block is expected.

(iv) Finally, a figure 8 conformation with $n$ turns per diblock may
be expected if $h_{n+1} \leq \Upsilon \leq h_n$.

\section{Brownian Dynamics Simulations}
\label{sec:sim}
As a compliment to the analytic treatment above we have conducted
Brownian dynamics simulations with the intention of confirming that
structures such as the figure 8 and toroids with $m > 1$ are realistic
polymer morphologies, and that their relative energies conform with our
theoretical predictions.

In our simulation model, poor solvent conditions were imposed by using
a Lennard Jones pair potential between like monomers ie $U_{AA} =
U_{BB} = 4\epsilon ((\sigma/r)^{12} - (\sigma/r)^6)$ , where $r$ is
the inter-monomer distance.  $\epsilon$ is the strength of the
interaction and is our unit of energy. $\sigma$ is roughly equal to
the equilibrium inter-bead distance and is therefore related to the
parameter $a$ in our analytic work. We chose $\sigma = 2$ for all
simulations. For simplicity in our analytic work we specified that
blocks of A type monomers and B type monomers cannot have lateral
contact.  Of course, such a rigid constraint cannot occur for a real
polymer and a more realistic situation is that monomers A and B
experience a repulsive interaction which offsets their solvent induced
attraction. We model this in our simulation by specifying the pairwise
excluded volume interaction between monomers A and B as, $U_{AB} = 4
\epsilon (\sigma/2r)^{12} $.

In order to account for the stiffness of the chain we used the
potential $U_{bend} = -\kappa^{\prime} cos\theta$ where $\theta$ is
the angle between successive bonds and $\kappa^{\prime}$ is the
bending constant.  $\kappa^{\prime}$ is related to the quantity
$\kappa$ used in analytic work through the relation $\kappa^{\prime} =
\kappa a$. 

Bonds between successive monomers were modeled using the potential
$U_{bond} = kTd^{-1}\ell^*(r/(d))$ which is equivalent to a Gaussian
chain entropic spring potential $3kTr^2/(2d^2)$ at short extensions
but increases rapidly for extensions close to or beyond the bond
length $d$. For simplicity, we chose a bond-length equal to the bead
diameter such that $\sigma \approx a = d = 2$. We calculated particle
motion by integrating the Langevin equation with a time-step of $\Delta
t = 0.002$ a temperature of $k_BT = 1$, a frictional coefficient
$\zeta = 1$ and random fluctuations $f(t)$ given by a Gaussian
distribution with mean $0$ and variance $\sqrt{2k_BT\Delta t}$.  All
simulations were carried out for a total of $25 \times 10^6$
time-steps.

\begin{figure}
\label{j39}
\scalebox{0.8}{\includegraphics*{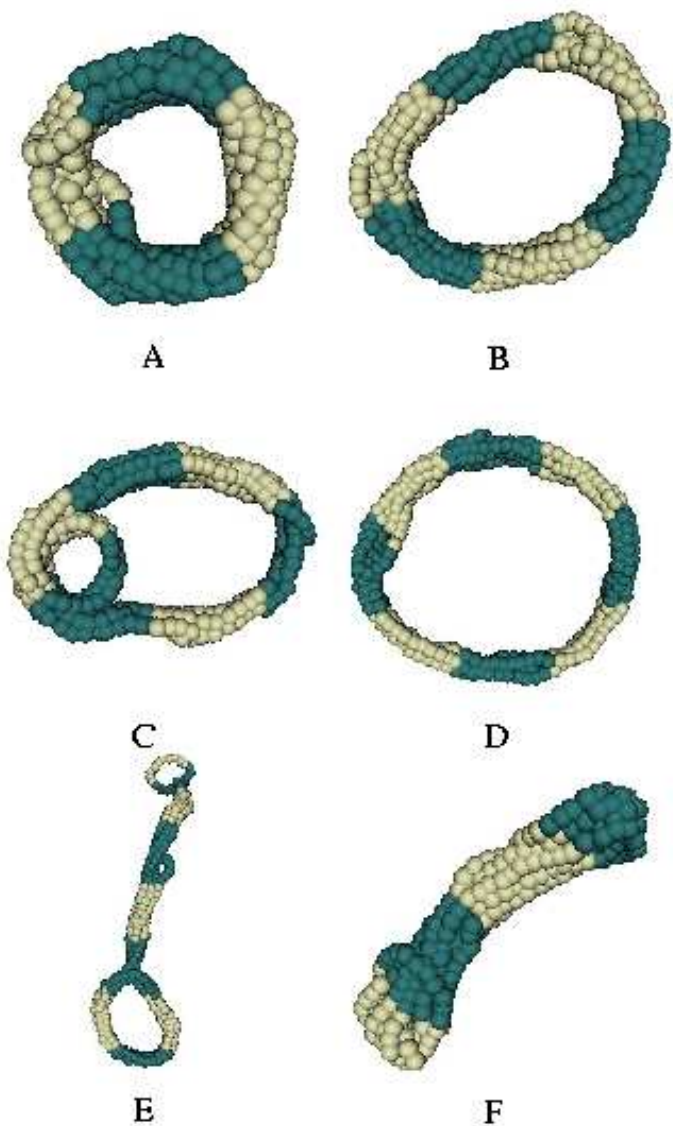}}
\caption{Steady-state collapsed configurations of semiflexible
copolymer where $N=20$, $L=400$, $\kappa = 15$.  Snapshots shown
illustrate the diversity of collapsed states and are labeled
alphabetically.  The relative internal energies of these states, $U (
 \epsilon ) $ and their frequency of occurrence $f$ as a percentage of
the $40$ replicate simulation runs are as follows. A: $U = 0 \pm 10$,
$f = 20\%$, B: $U=115 \pm 20$, $f = 10\%$, C: $U=30 \pm 20$, $f=20\%$,
D: $U=250 \pm 10$, $f=12.5\%$, E: $U=360$,$f=2.5\%$, F: $U=300 \pm
40$, $f=27.5\%$ . Overall values for $U$ and associated errors are
calculated from the set of all replicates that collapsed to a
particular configuration. Values for each of these simulations in turn
were taken over the final $1 \times 10^7$ time steps of the
simulation.}
\end{figure}

Since our intention was to observe striped toroidal structures with
different numbers of blocks per turn, $m$ we initially chose to
simulate a polymer with a large number of blocks $N = 20$.  The total
chain length was $400$ monomers, which gives $L=800$ and $l = L/N =
40$.  Stiffness was chosen to be relatively large $\kappa^{\prime} =
15$ so as to avoid globular configurations which are not accounted for
in our analytic model.  We conducted $40$ replicate runs in which a
polymer with the above parameters was quenched from an equilibrated
configuration in good solvent conditions to poor solvent via the
imposition of inter-monomer LJ potentials.  We found that after an
initial period of collapse, a steady state configuration was reached
which showed no change over at least the final half of the simulation
time. A wide variety of such configurations were observed, including
those with $m = 2, 3, 4$ as well as some previously unexpected
morphologies.  Since entropic effects are negligible in our poor
solvent system the free energies of each of these configurations are
roughly equal to their internal energies $U$.  Under this assumption
we found that the $m = 2$ configuration was the most stable of those
observed, with all other states including toroids with $m=3$ and $4$
all being higher in energy.

Close examination of the configurations in figure \ref{j39} reveals
several points of particular note.  Firstly, we see that the repulsion
between A and B type monomers effectively results in a complete
segregation of these blocks.  The end result is almost identical to
the constraint used in our analytic calculations that A and B blocks
cannot have lateral contact.  What is not predicted in our analytic
theory is that this segregation of A and B blocks leads to the
formation of a diverse range of metastable structures.  This occurs
because A and B type monomers cannot slide against one another, which
is an important equilibration mechanism for homopolymers.  As a
consequence, we find that structures with $m = 3$ and $m = 4$ cannot
readily transform to the more stable $m = 2$ type torus.  Instead they
remain as metastable states.  In addition to the striped toroidal
states we also observe other interesting metastable structures such as
a torus within a torus (figure \ref{j39} C) or a torus at the end of a
rod (figure \ref{j39} E) or a striped rod (figure \ref{j39} F).  In
all cases it seems likely that the A B segregation prevents a
transition to a state of lower free energy.
\begin{figure}
\label{j3}
\scalebox{0.8}{\includegraphics*{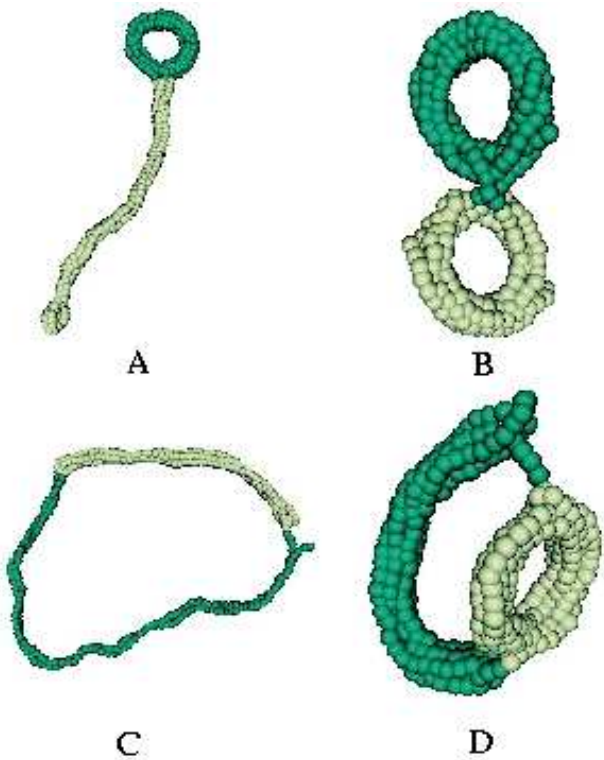}}
\caption{Steady-state collapsed configurations of semiflexible
copolymer where $N=2$, $L=800$, $\kappa = 15$ .  Snapshots shown
illustrate the diversity of collapsed states and are labeled
alphabetically.  The relative internal energies of these states, $U
(\epsilon) $ and their frequency of occurrence $f$ as a percentage of
the $40$ replicate simulation runs are as follows. A: $U = 70 \pm 10$,
$f = 37.5\%$, B: $U=0 \pm 10$, $f = 15\%$, C: $U=440 \pm 10$, $f=5\%$,
D: $U=140$, $f=2.5\%$.  Rods $U=220 \pm 10$,$f=35\%$ are not
shown. See figure \ref{j39} for details on calculation of averages and
errors}
\end{figure}

Although the simulation parameter $\epsilon$ is clearly related to the
surface tension $\gamma$, this relationship is very difficult to
obtain from a first principles calculation.  Instead, we can calibrate
$\gamma$ for our simulation system by using the internal energies of
the striped toroidal states.  The relative energies of these states
should be given by equation \ref{eqn5} so by fitting this equation to
values of $U$ and allowing for an unknown additive constant, we can
obtain a rough estimate for $\gamma$.  The value of $\gamma$ so
obtained was $\gamma \approx 0.4$.  By substituting this value of
$\gamma$ back into equation \ref{eqn5} we find that the energy of the
$m=1$ torus for our simulated polymer is predicted to be $U \approx
800 \epsilon$. Such a high predicted free energy is consistent with
the fact that we never observed the formation of $m=1$ toroids for the
$N=20$ polymer.  Our value of $\gamma$ can also be used to predict an
appropriate set of parameters for the observation of figure 8
structures. In making such a prediction we keep the total polymer
length $L$ constant, such that $l = L/N$, and vary $N$.  Keeping all
other parameters constant we predict that figure 8 configurations
should be observed for $N < 6$.  Taking a conservative approach to
this prediction we simulated a polymer with $N=2$ using the same
methods as were used for the $N=20$ case.

In figure \ref{j3} we present snapshots of steady state collapsed
configurations of the $N=2$ polymer.  The most stable observed
configuration in this case is the figure 8, however as with the $N=20$
polymer many metastable configurations were also observed. Indeed the
most commonly observed shape was that of a single torus attached to a
rod.  Other metastable states included complete rods as well as an $m
= 1$ toroid and a strange configuration in which a toroid exists
between two ends of a bent rod.  Although this last configuration is
rare it serves to illustrate the diversity of metastable states which
occur when A and B type blocks repel one another.  Provided the A B
repulsion is sufficiently strong we would expect that many of the the
observed metastable states should be quite long lived.

\section{Concluding Remarks}
Semiflexible copolymers exhibit the potential for formation of a wide
variety of single molecule compact structures in poor solvent.  Even
the very simple alternating A B block case discussed here forms a
range of torii with discrete sizes as well as the unusual figure 8
structure.  Aside from this variety of structures one of the key
features of these copolymer collapsed states is their well defined
size which must be an integer multiple of the block length.  Although
such structures are yet to be realized we expect that they could have
great utility as tools in single molecule experiments, where for
example a molecule is pulled through a precisely defined pore ( eg
torus ) \cite{jiali_etal:2003}.

\bibliographystyle{unsrt}
\bibliography{main}

\end{document}